\documentclass[floatfix,reprint,amsmath,amssymb,aps,pra]{revtex4-1}
\usepackage{times,mathptmx}
\usepackage{graphicx}
\usepackage[english]{babel}
\usepackage{epstopdf}
\usepackage{longtable}
\usepackage{xcolor}

\begin{document}

\title{CPA-lasing associated with the quasibound states in the continuum in asymmetric non-Hermitian structures}
%\title{Asymmetry-induced transition from the quasibound state in the continuum to the CPA-lasing in the non-Hermitian multilayer structures}
\author{Denis~V.~Novitsky$^{1}$}
\email{dvnovitsky@gmail.com}
\author{Adi\`a~Can\'os~Valero$^{2}$}
\author{Alexander Krotov$^{2,3}$}
\author{Toms Salgals$^2$}
\author{Alexander~S.~Shalin$^{2,4}$}
\author{Andrey~V.~Novitsky$^{5,6}$}

\affiliation{$^1$B.~I.~Stepanov Institute of Physics, National Academy of Sciences of Belarus, Nezavisimosti Avenue 68, 220072 Minsk, Belarus \\ 
$^2$Riga Technical University, Institute of Telecommunications, Azenes st. 12, 1048 Riga, Latvia \\
$^3$Saint Petersburg Electrotechnical University “LETI” (ETU),  Prof. Popova Street 5, 197376 St. Petersburg, Russia \\
$^4$Kotel’nikov Institute of Radio Engineering and Electronics, Russian Academy of Sciences (Ulyanovsk branch), Goncharova Str. 48, 432000 Ulyanovsk, Russia \\
$^5$Department of Theoretical Physics and Astrophysics, Belarusian State University, Nezavisimosti Avenue 4, 220030 Minsk, Belarus \\
$^6$ITMO University, Kronverksky Prospekt 49, 197101 St. Petersburg, Russia}

\date{\today}

\begin{abstract}
Non-Hermitian photonic systems with loss and gain attract much attention due to their exceptional abilities in molding the flow of light. Introducing asymmetry to the $\mathcal{PT}$-symmetric system with perfectly balanced loss and gain, we reveal the mechanism of transition from the quasibound state in the continuum (quasi-BIC) to the simultaneous coherent perfect absorption (CPA) and lasing in a layered structure comprising epsilon-near-zero (ENZ) media. Two types of asymmetry (geometric and non-Hermitian) are analyzed with the scattering matrix technique. The effect of the CPA-lasing associated with the quasi-BIC is characterized with the unusual linear dependence of the quality factor on the inverse of asymmetry parameter. Moreover, the counter-intuitive loss-induced-lasing-like behavior is found at the CPA-lasing point under the non-Hermitian asymmetry. The reported features of non-Hermitian structures are perspective for sensing and lasing applications.
\end{abstract}

\maketitle

\section{Introduction}

Since the pioneering work of $\mathcal{PT}$-symmetric quantum mechanics \cite{Bender1998}, non-Hermitian physics became an extremely wide field of research concerning not only quantum effects but also branches of classical physics including photonics, mechanics, electrical engineering, and even biophysics \cite{Ashida2020}. Principles of the non-Hermitian photonics have stimulated an especially fruitful design of novel optical systems with loss and gain elements. $\mathcal{PT}$-symmetric structures with balanced loss and gain as a particular case of non-Hermitian systems have attracted most attention due to their ability to implement basic effects of both $\mathcal{PT}$-symmetric response and symmetry breaking \cite{Zyablovsky2014, Feng2017, El-Ganainy2018, Ozdemir2019}. Applications of $\mathcal{PT}$ symmetry in optics and photonics are diverse and include sensing with enhanced sensitivity \cite{Chen2017, Hodaei2017, Wiersig2020}, slowing of light \cite{Goldzak2018}, effective single-mode lasing \cite{Feng2014-2, Hodaei2014, Qi2019, Perriere2021}, coherent perfect absorption (CPA) \cite{Longhi2010, Wong2016, Novitsky2019}, topological-protection of surface states \cite{Parto2021}, and even training of optical neural networks \cite{Deng2021}.

There are two basic geometries widely employed in the studies on non-Hermitian photonics \cite{Zyablovsky2014}. Longitudinal geometry is used in single-mode \cite{Ruter2010} and multimode coupled waveguides \cite{Hlushchenko2020, Hlushchenko2021a, Hlushchenko2021b}, one- \cite{Regensburger2012} and two-dimensional photonic lattices \cite{Kremer2019}, and coupled microcavities \cite{Wen2018}. The most popular transverse geometry is a multilayer structure with alternating loss and gain media. Being a non-Hermitian generalization of the photonic crystal concept, such a multilayer attracts much attention due to its simplicity for analysis and availability for unusual optical responses, such as anisotropic transmission resonances \cite{Ge2012}, resonant scattering \cite{Shramkova2016}, nonlinear saturation effects \cite{Witonski2017, Novitsky2018, Shramkova2019, Shramkova2021}, nonlocality \cite{Novitsky2019a}, pulse-propagation effects \cite{Tsvetkov2019, Shestakov2021}, effects of disorder \cite{Novitsky2021a}, etc. From the more general perspective, many of these effects can be treated as ``anomalies'' in light scattering \cite{Krasnok2019, Krasnok2020} being described by means of scattering matrix technique \cite{Ge2012, Novitsky2020}. The features of light scattering on dielectric structures were deeply studied in recent years \cite{Terekhov2019, Barhom2019, Valero2020, Terekhov2020, Shamkhi2019}.

A special type of ``anomaly'' is a singular optical response in media with permittivity close to zero, which are called epsilon-near-zero (ENZ) media. The ENZ media demonstrate a number of unique properties such as wavelength expansion, field enhancement, light tunneling, light velocity manipulation, and strong nonlinear and nonlocal effects \cite{Liberal2017, Kinsey2019, Reshef2019}. Here we are interested in the so-called bound states in the continuum (BICs) -- trapped modes of open cavities with perfect localization of radiation and resonances of infinite quality ($Q$) factor \cite{Hsu2016, Azzam2021, Sadreev2021}. ENZ-containing structures can support BICs both in waveguide \cite{Liu2021} and layered geometries \cite{Monticone2018, Duggan2019, Sakotic2020, Sakotic2021, Castaldi2021}.

In this paper, we study a non-Hermitian generalization of the ENZ-containing layered structures supporting bound states in the continuum. It has been recently shown \cite{Novitsky2021} that $\mathcal{PT}$-symmetry breaking in the structures with balanced loss and gain results in transformation of a BIC into quasi-BIC, which is the resonance with the finite $Q$ factor. Here, we make a further step introducing asymmetry in the distribution of loss and gain. The asymmetry (either in the thicknesses of layers or in the non-Hermiticity value itself, see Fig. \ref{fig1}) serves as an additional degree of freedom for transferring from the quasi-BIC to another ``anomaly'' effect -- the CPA-lasing point -- as is demonstrated using analysis of the scattering-matrix poles and zeros. The CPA-lasing in this case is directly associated with the quasi-BIC providing strong light amplification, linear dependence of the $Q$ factor on the inverse asymmetry parameter, and loss-induced-like lasing effect.

\section{Geometric asymmetry}

\begin{figure}[t!]
\includegraphics[scale=0.45, clip=]{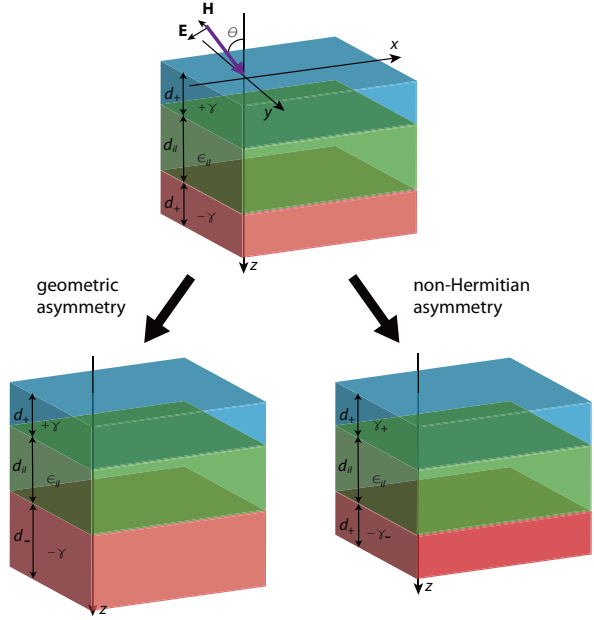}
\caption{\label{fig1} Schematic of an asymmetric trilayer consisting of the outer layers with loss and gain ENZ media, respectively, and the dielectric spacer. The outer layers are initially of the same thicknesses $d_+$ and non-Hermiticity parameters $\gamma$. Asymmetry has either geometric ($d_-=\alpha d_+$) or non-Hermitian ($\gamma_-=\beta \gamma_+$) origin as shown in the lower part of the figure. The spacer has the thickness $d_{il}=10 d_+$ and permittivity $\varepsilon_{il}=5$.}
\end{figure}

We start with the non-Hermitian trilayer system having outer loss and gain layers of different thicknesses and the spacer (interlayer) between them (Fig. \ref{fig1}). This case is characterized by violation of the gain-loss balance due to dimensions and, therefore, is called the geometric asymmetry. To be consistent with Ref. \cite{Novitsky2021}, we take the permittivity of the loss and gain media $\varepsilon_\pm (\omega)=1 \pm i \gamma -\omega^2_p/\omega^2$, where $\omega_p$ is the plasma frequency and $\gamma$ is the non-Hermiticity magnitude. Exploitation of the classical Drude-Lorentz model does not spoil the conclusions we make further. The loss layer has the thickness $d_+=\lambda_p/2 \pi$ (i.e., $\omega_p d_+/c=1$, where $c$ is the speed of light), whereas the gain layer has different thickness $d_-=\alpha d_+$, where $\alpha$ is the geometric asymmetry parameter. The parameters of the spacer are $d_{il}=10 d_+$ and $\varepsilon_{il}=5$.

\begin{figure}[t!]
\includegraphics[scale=0.95, clip=]{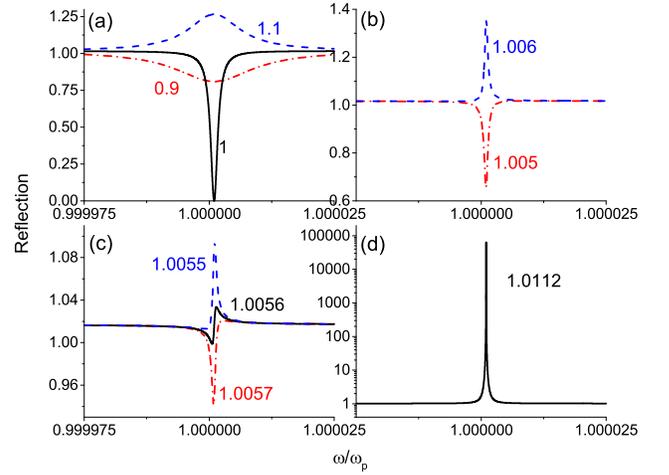}
\caption{\label{fig2} Spectra of reflection for different geometric asymmetry parameters $\alpha$ marked as numbers near the curves. The angle of incidence is $\theta=23.881^{\circ}$, the non-Hermiticity magnitude $\gamma=0.001$.}
\end{figure}

In our previous paper \cite{Novitsky2021}, the symmetric case ($\alpha=1$) has been already studied. In the Hermitian limit ($\gamma=0$), the symmetric system possesses an unobservable infinitely narrow resonance (BIC) due to destructive interference of the Fabry-Perot mode and the volume plasmon excited by TM-polarized waves at the plasma frequency. When $\gamma > 0$, the non-Hermiticity drives the transition of perfect BIC to the quasi-BIC through the mechanism of $\mathcal{PT}$ symmetry breaking. This quasi-BIC, which can be reached at the incidence angle $\theta_{BIC} = \arcsin \sqrt{\varepsilon_{il}- \left( \frac{\pi c n}{\omega_p d_{il}} \right)^2}$ ($n$ is the integer), shows up in spectra as a finite-width resonance, see the solid line in Fig. \ref{fig2}(a).

Let us fix the angle of incidence ($\theta=\theta_{BIC} \approx 23.881^{\circ}$ at $n=7$) and the non-Hermiticity magnitude ($\gamma=0.001$) and start changing the geometric asymmetry parameter $\alpha$. Reflection spectra calculated within the transfer-matrix method \cite{Novitsky2021} are shown in Fig. \ref{fig2}. For $\alpha<1$, the thickness of the gain layer gets smaller than the thickness of the loss layer, $d_- < d_+$, so that the quasi-BIC resonance dip just becomes broader and shallower [see the line at $\alpha=0.9$ in Fig. \ref{fig2}(a)]. The case of $\alpha>1$, when $d_- > d_+$, is much more interesting due to its richer physics connected to competition between loss and gain. As shown in Fig. \ref{fig2}(a), there is a broad resonance peak at $\alpha=1.1$, so that the transfer from the dip to the peak happens somewhere between $\alpha=1$ and $1.1$. Analysis shows that this transition happens at $\alpha_t \approx 1.0056$, when the system contains slightly more gain than loss. There is no paradox that $\alpha_t \neq 1$, since for $0 < \alpha < \alpha_t$ the gain results in the overall (wideband) reflection with intensity coefficient $R>1$, while the dip appears on this background. The resonance keeps symmetric line quite close to $\alpha_t$ [Fig. \ref{fig2}(b)] and becomes asymmetric only in the very vicinity of the transition. The line at $\alpha_t$ clearly has the Fano profile [Fig. \ref{fig2}(c)] stemming from the interplay between wide Fabry-Perot and narrow plasmonic resonances. Finally, close to some intermediate asymmetry parameter $\alpha_0 = 1.0112$, the reflection peak reaches maximum [Fig. \ref{fig2}(d)] and then gradually diminishes. The transition point $\alpha_t$ corresponds to the middle between the deepest dip at $\alpha=1$ and highest peak at $\alpha=1.0112$ in agreement with expected linearity of the response on small perturbation.

\begin{figure}[t!]
\includegraphics[scale=0.95, clip=]{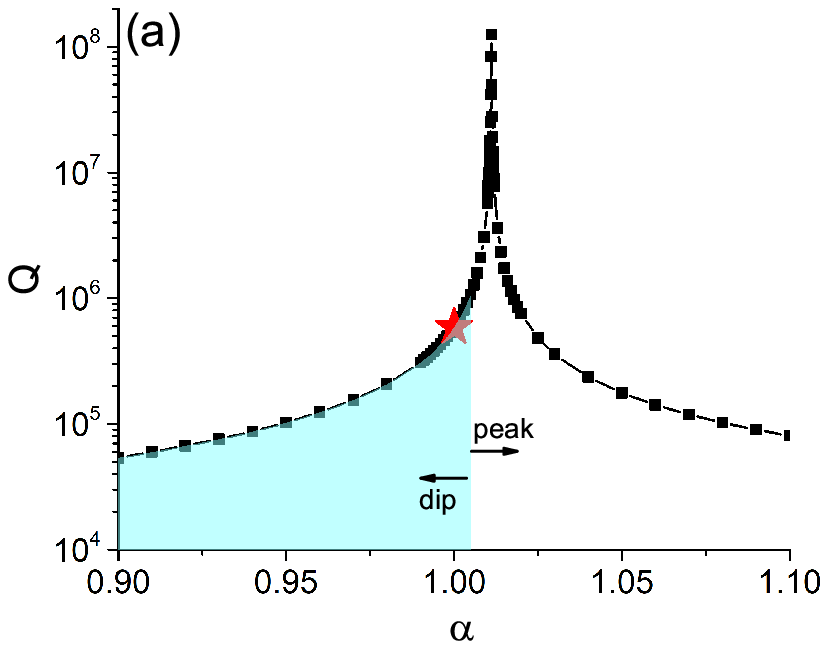} \includegraphics[scale=0.95, clip=]{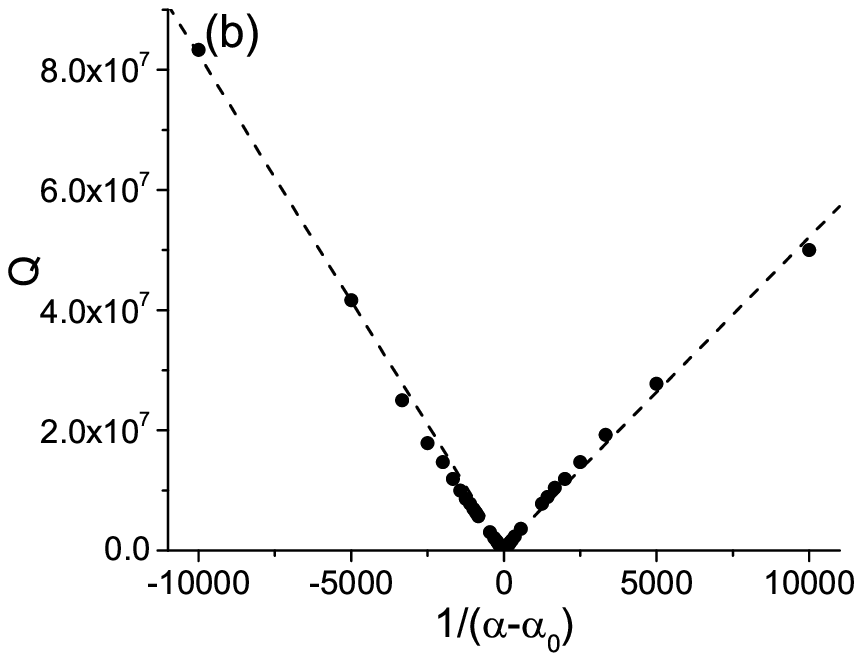}
\caption{\label{fig3} Dependence of the $Q$ factor on the geometric asymmetry parameter $\alpha$ (a) and on the value $1/(\alpha-\alpha_0)$. The angle of incidence is $\theta=23.881^{\circ}$, the non-Hermiticity magnitude $\gamma=0.001$. The red star shows the quasi-BIC position at $\alpha=1$, whereas the filled area marks the transition between the dip and peak.}
\end{figure}

As a quantitative characteristic of the observed resonances, we calculate the quality ($Q$) factor. Due to the symmetry of lineshapes outside the immediate vicinity of the transition from the dip to peak, it can be estimated with the simple relation $Q=\omega_0/\Delta \omega$, where $\omega_0$ is the frequency of the resonance peak or dip and $\Delta \omega$ is the full width of the resonance. Such a consideration is in accordance with the modern treatment of non-Hermitian resonances using the quasi-normal modes characterized by complex frequencies \cite{Wu2021}. Dependence of the $Q$ factor on the parameter $\alpha$ calculated with this expression is shown in Fig. \ref{fig3}. One can see that the powerful peak at $\alpha \approx \alpha_0$ in Fig. \ref{fig2}(d) corresponds to the sharp increase of the $Q$ factor exceeding $10^8$. This amplification of the reflection and transmission are linked to strong localization of radiation inside the structure. 

Figure \ref{fig3}(b) surprisingly shows that the $Q$ factor is inversely proportional to the geometric asymmetry parameter $\alpha$. This result is in contrast to what is observed in usual BICs, which demonstrate the inverse proportionality of $Q$ to the square of the asymmetry factor \cite{Koshelev2018}. On the other hand, the recently introduced non-Hermitian BICs (so-called \textit{pt}-BICs) reported in Ref. \cite{Song2020} have precisely such an inverse behavior of $Q$ as a function of the longitudinal wavenumber playing the role of the asymmetry parameter. The fundamental difference with that case is that our system is not $\mathcal{PT}$ symmetric due to the asymmetry between loss and gain layers, so that the maximal $Q$ factor is reached at some $\alpha_0>1$.

\begin{figure}[t!]
\includegraphics[scale=1., clip=]{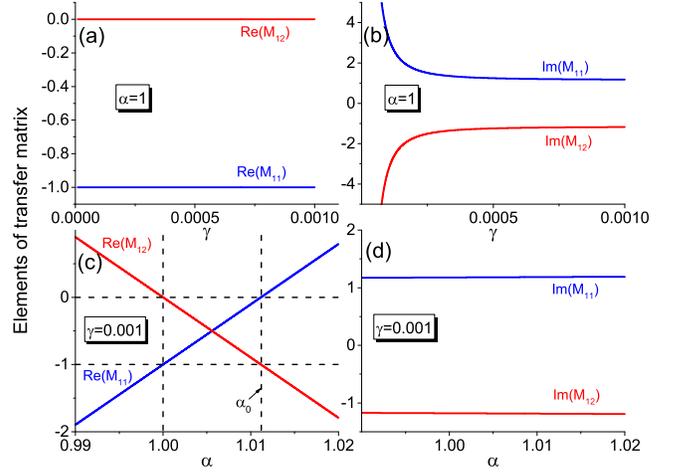} 
\caption{\label{fig4} Dependence of the transfer-matrix elements on (a)-(b) the non-Hermiticity parameter $\gamma$ for the symmetric structure and (c)-(d) the asymmetry parameter $\alpha$ for $\gamma=0.001$. The angle of incidence is $\theta=23.881^{\circ}$, the frequency is $\omega=\omega_p$.}
\end{figure}

What is the nature of the $Q$ factor enhancement at $\alpha_0$? To address this question, let us consider poles and zeros of the scattering matrix of our system. The scattering matrix gives the equivalent results as the effective Hamiltonian approach \cite{Novitsky2020}, but is more convenient to analyze the multilayer structures. In general, the scattering matrix of a multilayered structure has the form $\hat S = \left(
\begin{array}{cc} t & r_R \\ r_L & t \end{array} \right)
\label{scat}$ \cite{Novitsky2020}. Here $t=1/M_{11}$ is the transmission coefficient and $r_L=M_{21}/M_{11}$ and $r_R=-M_{12}/M_{11}$ are the reflection coefficients for the left- and right-incident waves which can be determined through the corresponding elements of the transfer matrix $M$ \cite{Novotny, Novitsky2021}. A pole of the scattering matrix can be determined from the condition of the mode self-excitation arising at $t=\infty$ or, equivalently, $M_{11}=0$. According to Ref. \cite{Krasnok2019}, the condition for a scattering matrix zero is $t \pm r=0$ resulting in $M_{12}= \pm 1$ or $M_{21}= \pm 1$. The conditions for poles and zeros are generally fulfilled at complex frequencies. A BIC appears in Hermitian systems when pole and zero meet at the real axis of the complex-frequency plane \cite{Krasnok2019}. One can directly confirm that this happens in the symmetric trilayer at $\gamma=0$, $\omega=\omega_p$ and $\theta=\theta_{BIC}$, where the elements of the transfer matrix become real with $M_{11}=0$, $|M_{12}| = |M_{21}| = 1$. Introduction of non-Hermiticity breaks this exact BIC down as reported in Ref. \cite{Novitsky2021}. Figures \ref{fig4}(a) and \ref{fig4}(b) show how this looks like from the viewpoint of poles and zeros. One can see that both elements of the transfer matrix are complex for $\gamma \neq 0$, with real parts $\mathrm{Re} M_{11}=-1 \neq 0$ and $\mathrm{Re} M_{12}=0 \neq \pm 1$, so that the condition for the coalescence of pole and zero is not fulfilled anymore (since $M_{21}=M_{12}^\ast$, further we consider only $M_{12}$).

Influence of the geometric asymmetry on the components of the transfer matrix is shown in Figs. \ref{fig4}(c) and \ref{fig4}(d). Whereas their imaginary parts change only slightly, the real parts demonstrate clear linear dependence on $\alpha$, so that $\mathrm{Re} M_{11} \approx 0$ and $\mathrm{Re} M_{12} \approx -1$ at $\alpha=\alpha_0$. Although the imaginary parts do not vanish, the values of the real parts allow us to suggest that we effectively reach both the pole and zero at $\alpha=\alpha_0$. Nonzero imaginary parts only limit possible values of the $Q$ factor making it finite. Thus, the asymmetry can be used to compensate the mismatch between pole and zero induced by non-Hermiticity. Moreover, one can estimate the change of the transfer-matrix elements with $\alpha$ by using the analytical expression for $M$. Since this expression is very cumbersome, it is more convenient for illustration to substitute all the parameters, except $\alpha$, and obtain the numerical dependence $M(\alpha)$. For example, we result in $M_{11}(\alpha) \approx (-93.17 + 0.58 i) \cosh [(0.40 + 0.0012 i) \alpha] + (240.18 + 0.67 i) \sinh [(0.40 + 0.0012 i) \alpha] \approx -(1 - 1.18 i) + (89.62 + 0.66 i) (\alpha - 1)$, which gives a very good linear approximation for $M_{11}(\alpha)$ shown in Fig. \ref{fig4}(c) and allows to estimate $\alpha_0$.

It is known that in non-Hermitian systems the coincidence of pole and zero heralds an intriguing effect of simultaneous coherent perfect absorption (CPA) and lasing \cite{Krasnok2019}. In our case, the lasing can be immediately associated with the sharp growth of the $Q$ factor at $\alpha=\alpha_0$. To further corroborate this interpretation, in Appendix A, we demonstrate the CPA effect in our structure for the same conditions as lasing.

Thus, the peak of the $Q$ factor reported above has the nature of CPA-lasing point. However, appearance of this point in the asymmetric structure is closely related to the quasi-BIC in the symmetric one. The sharp increase of the $Q$ factor and the lasing at $\alpha_0$ are achievable for the gain layer which is only $1.12 \%$ thicker than the loss layer. Such a small difference in thickness matters, because light intensity is localized under the conditions close to the quasi-BIC resonance. Therefore, we call the observed effect \textit{the CPA-lasing associated with the quasi-BIC}.

\begin{figure}[t!]
\includegraphics[scale=1., clip=]{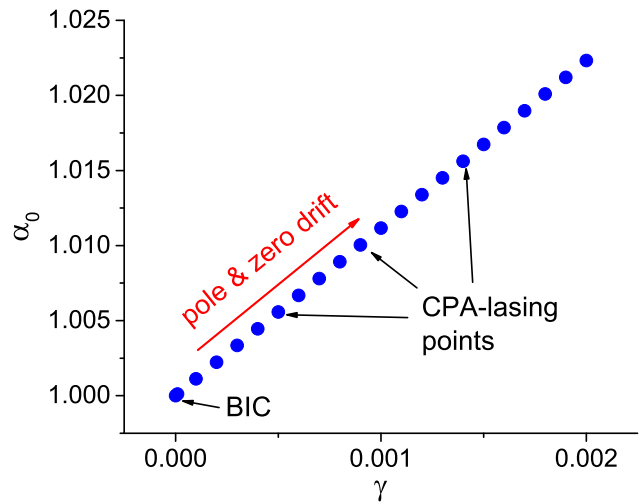} 
\caption{\label{fig5} The change of the geometric asymmetry parameter $\alpha_0$ corresponding to the pole and zero coincidence with the non-Hermiticity parameter $\gamma$.}
\end{figure}

The previous consideration was performed at the specific value of loss and gain parameter $\gamma$. The same approach based on the pole and zero dynamics and allowing to estimate the CPA-lasing asymmetry parameter $\alpha_0$ can be applied to any level of non-Hermiticity. As a result, we obtain the dependence $\alpha_0 (\gamma)$ shown in Fig. \ref{fig5}. The BIC is observed in the Hermitian case ($\gamma=0$) and in the symmetric structure with $\alpha_0=1$. Increasing $\gamma$ leads to the drift of pole and zero coincidence to the nonunitary asymmetry, $\alpha_0>1$. As a result, we obtain the line of CPA-lasing points in the plane $(\gamma, \alpha_0)$. It is interesting to note that this line is clearly straight indicating the linear dependence between non-Hermiticity and asymmetry in the analyzed system.

If the condition for the incidence angle ($\theta=\theta_{BIC}$) is relaxed (for example, we deal with the normal incidence of light), the much thicker gain medium should be used to reach the lasing. In that case, however, lasing is due to the pole dynamics only, in contrast to our case. Thus, the use of the CPA-lasing point associated with the quasi-BIC provides a different mechanism and allows to optimize the lasing structures, in particular lower the lasing threshold.

The CPA-lasing phenomenon associated with the quasi-BIC can be compared with the similar effect reported by Song et al. \cite{Song2020}. In that paper, it is shown that a $\mathcal{PT}$-symmetric perturbation splits the BIC into the \textit{pt}-BIC and lasing threshold modes. Song et al.'s \textit{pt}-BIC is characterized by the $Q$ factor having the linear dependence on the inverse asymmetry parameter in contrast to the usual for BICs inverse dependence on the squared asymmetry parameter. In our case, the dependence is also linear, but the situation is fundamentally different: the asymmetry perturbation violates existing $\mathcal{PT}$ symmetry of the loss-gain distribution and transforms the quasi-BIC into the CPA-lasing point. We emphasize the role of asymmetry which distinguishes our results also from the CPA-lasing arising from the BIC under the $\mathcal{PT}$-symmetric perturbation in electronic circuits \cite{Sakotic2022}.

\section{Non-Hermitian asymmetry}

\begin{figure}[t!]
\includegraphics[scale=1., clip=]{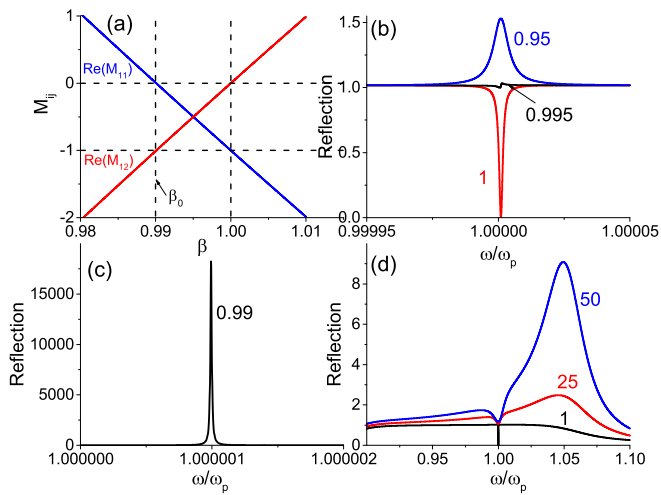}
\caption{\label{fig6} (a) Dependence of the transfer-matrix elements on the asymmetry parameter $\beta$ at the plasma frequency. (b)-(d) Spectra of reflection for different non-Hermitian asymmetry parameters $\beta$ marked as numbers near the curves. The angle of incidence is $\theta=23.881^{\circ}$, the loss magnitude $\gamma_+=0.001$.}
\end{figure}

\begin{figure}[t!]
\includegraphics[scale=1., clip=]{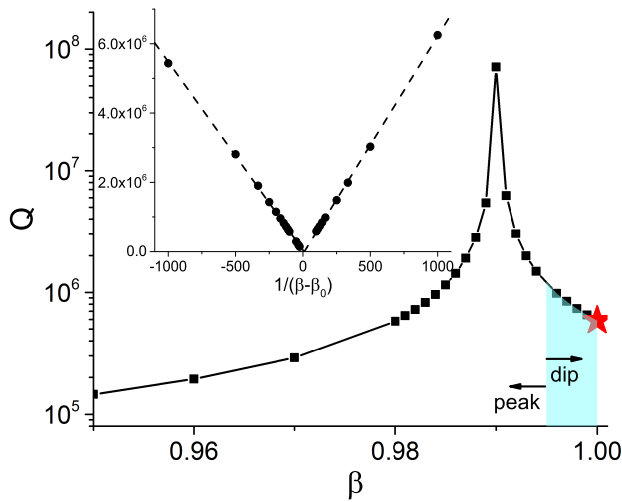}
\caption{\label{fig7} Dependence of the $Q$ factor on the non-Hermitian asymmetry parameter $\beta$ and on the value $1/(\beta-\beta_0)$ (inset). The angle of incidence is $\theta=23.881^{\circ}$, the loss magnitude $\gamma_+=0.001$. The red star shows the quasi-BIC position at $\alpha=1$, whereas the filled area marks the transition between the dip and peak.}
\end{figure}

In this section, we consider a geometrically symmetric ($d_+=d_-$) trilayer structure with violated balance between loss and gain. We introduce another asymmetry parameter $\beta=\gamma_-/\gamma_+$, where $\gamma_+$ and $\gamma_-$ are the loss and gain magnitudes, respectively. The asymmetry realized through parameter $\beta$ can be called non-Hermitian asymmetry. Condition $\beta>1$ implies dominating impact of gain in system's response. One can expect that $\beta>1$ has a similar effect as the increase of the gain layer thickness ($\alpha>1$) discussed above. However, this is not the case. 

As shown in Fig. \ref{fig6}(a), the elements of the transfer matrix approach the CPA-lasing condition (simultaneous $\mathrm{Re} M_{11} \approx 0$ and $\mathrm{Re} M_{12} \approx -1$) at $\beta=0.99$, i.e., in the overall lossy system. This results in a surprising increase of reflection with decreasing $\beta$ as corroborated with the spectra shown in Fig. \ref{fig6}(b) with the dip-to-peak transition at $\beta_t \approx 0.995$. The CPA-lasing condition at $\beta=0.99$ corresponds to the extremely amplified reflection [Fig. \ref{fig6}(c)] and the strongly enhanced $Q$ factor (Fig. \ref{fig7}). As in the case of geometric asymmetry, the $Q$ factor linearly depends on inverse of the non-Hermitian asymmetry parameter (see the inset in Fig. \ref{fig7}), although the peak value of $Q$ is somewhat lower due to the lossy character of the structure. 

On the contrary, for the systems with overall gain, we do not see any significant increase of the reflected (and transmitted) radiation in vicinity of the plasma frequency even for as great asymmetry parameter as $\beta=50$ [Fig. \ref{fig6}(d)]. Formation of the dip for $\beta>1$ can be explained as the quasi-BIC in the symmetric case ($\beta=1$) broadened and elevated by the overall gain in the asymmetric structure. Amplification of the reflection (as well as transmission) happens mostly above the plasma frequency where the outer layers are dielectric-like. We should also emphasize the narrowband nature of CPA-lasing [Fig. \ref{fig6}(c)] as opposed to the strongly wideband usual amplification [Fig. \ref{fig6}(d)] which can be of interest for possible applications.

Our analysis demonstrates that the CPA-lasing point associated with the quasi-BIC can be reached using the non-Hermitian asymmetry either. Its behavior is caused by the intriguing interplay of loss and gain and provides one more intriguing loss-induced effect in addition to the loss-induced transparency \cite{Guo2009} and loss-induced lasing \cite{Peng2014} in non-Hermitian systems. The difference is that we use the asymmetry as a driver for such an unusual response possible due to the quasi-BIC proximity.

The above considerations have been performed at the plasma frequency where the true BIC exists in the Hermitian limit and the non-Hermiticity-induced quasi-BIC resonance has a symmetric lineshape. At the nearby frequencies, the asymmetric Fano resonances appear at the angles different from $\theta_{BIC}$ and can be used to realize the CPA-lasing-like effect in asymmetric structures as well. The corresponding examples are discussed in Appendix B.

\section{Conclusion}

To sum up, we have introduced the concept of the CPA-lasing associated with the quasi-BIC unveiled in asymmetric non-Hermitian ENZ-containing layered structures. One can say that asymmetry in these structures supports transformation of the quasi-BIC resonance into the CPA-lasing resonance. The cases of different thicknesses of loss and gain layers (geometric asymmetry) and unequal levels of loss and gain (non-Hermitian asymmetry) have been studied. The effects of asymmetry have been analyzed in the framework of poles and zeros of the scattering matrix revealing intriguing features. We have determined a CPA-lasing point of merging pole and zero characterized by strong amplification of the outcoming intensity and sharp increase of the $Q$ factor associated with the nearby quasi-BIC. We would like to highlight two important results found out in this paper. First, the $Q$ factor has unusual inverse linear dependence on the asymmetry parameter. Second, the counter-intuitive loss-induced amplification is discovered in the system with non-Hermitian asymmetry at the plasma frequency. We believe that the results reported here are of general interest for non-Hermitian photonics and can be extended to two- and three-dimensional systems. The CPA-lasing effect associated with the quasi-BIC is envisaged to be demanded in laser and nonlinear-optics systems based on strong light enhancement.

\acknowledgements{The work was supported by the Belarusian Republican Foundation for Fundamental Research (Projects No. F20R-158) and the Russian Foundation for Basic Research (Project No. 20-52-00031). The loss-induced lasing-like behaviour investigation has been partially supported by the Russian Science Foundation (Grant No. 21-12-00151). The poles and zeros calculations have been partially supported by the Russian Science Foundation (Grant No. 21-12-00383).}

\appendix

\section{CPA-lasing effect via counter-propagating waves}

\begin{figure}[t!]
\includegraphics[scale=0.95, clip=]{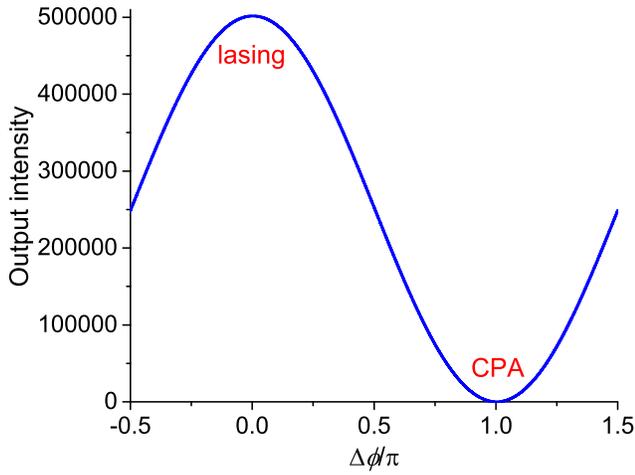} 
\caption{\label{figS1} The output intensity as a function of the phase difference $\Delta \phi$ (in units of $\pi$) between the two counter-propagating waves. The non-Hermiticity parameter is $\gamma=0.001$, the angle of incidence is $\theta=23.881^{\circ}$, the geometric asymmetry parameter $\alpha=1.0112$, the frequency is $\omega=1.000001 \omega_p$ (corrected to take the small shift of the resonance due to the asymmetry).}
\end{figure}

To directly demonstrate the CPA-lasing effect in our system, we consider two counter-propagating waves as in Refs. \cite{Wong2016, Novitsky2019}. Depending on the phase difference $\Delta \phi$ between these waves, we expect observing either CPA or lasing. Calculations of the output intensity (sum of reflected and transmitted intensities from all interfaces) demonstrated in Fig. \ref{figS1} validate these expectations. Indeed, for $\Delta \phi = 0$, the output intensity is maximal and is five orders of magnitude larger than the input one (lasing). To the contrary, $\Delta \phi = \pi$ corresponds to the output intensity close to zero (CPA). Notice that such values of the phase difference $\Delta \phi$ associated with lasing and CPA are due to presence of the ENZ medium \cite{Bai2016}. To the contrary, in the case of usual (positive) permittivities, $\Delta \phi = \pi/2$ and $\Delta \phi = -\pi/2$ are used for CPA and lasing observation \cite{Novitsky2019}.

\section{CPA-lasing at Fano resonances}

\begin{figure}[b!]
\includegraphics[scale=1., clip=]{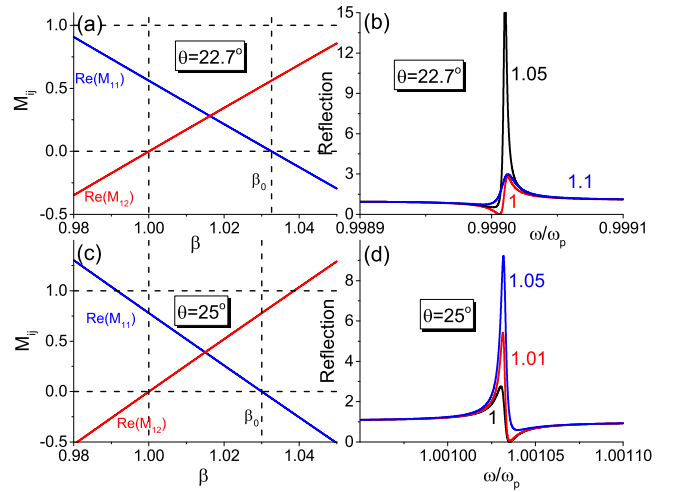}
\caption{\label{figS2} (a), (c) Dependence of the transfer-matrix elements on the asymmetry parameter $\beta$ at the angles of incidence $\theta=22.7^{\circ}$ (at frequency $\omega=0.999\omega$) and $\theta=25.0^{\circ}$ (at frequency $\omega=1.001\omega$) and (b), (d) the corresponding spectra of reflection for different non-Hermitian asymmetry parameters $\beta$ marked as numbers near the curves. The loss magnitude $\gamma_+=0.001$.}
\end{figure}

An example regarding the non-Hermitian asymmetry at various angles is shown in Fig. \ref{figS2}. We take two angles: the first one $\theta=22.7^{\circ}$ is below $\theta_{BIC}$, while the second angle $\theta=25.0^{\circ}$ is above it. The resonance frequency position shift to the values $\omega=0.999\omega$ and $\omega=1.001\omega$, respectively. One can see that the pole condition $\mathrm{Re} M_{11} \approx 0$ is achieved at $\beta_0=1.03263$ [Fig. \ref{figS2}(a)] and $\beta_0=1.03$ [Fig. \ref{figS2}(c)], while the zero condition ($\mathrm{Re} M_{12} \approx 1$) is not reached at these $\beta_0$. However, even such an incomplete CPA-lasing condition is enough for strong amplification of the reflection near $\beta_0$ as seen in Figs. \ref{figS2}(b) and \ref{figS2}(d). 

The amplification confirms the same nature of the effect at any frequency and its tolerance to the parameters variation. Nevertheless, the case of the plasma frequency can be distinguished for two reasons. First, there is a pronounced asymmetry of the Fano spectra profiles in Fig. \ref{figS2} as compared to the symmetric profiles in Fig. 6 of the main text. Second, $\beta_0>1$ both below and above $\theta_{BIC}$ being opposed to the counter-intuitive case of $\beta_0<1$ at the incidence angle $\theta=\theta_{BIC}$. These facts confirm the specific nature and unusual features of response close to the BIC at the plasma frequency.

\end{document}